\documentstyle[preprint,aps,prd,epsfig]{revtex}
\def\lsim{\raise0.3ex\hbox{$<$\kern-0.75em\raise-1.1ex\hbox{$\sim$}}}

\preprint{          Preprint numbers: \ \
          BI-TP 98/15 \ \
          UUHEP 98/3 \ \
}
\begin{document}
\draft

\title{String Breaking in Lattice Quantum Chromodynamics}

\author{Carleton DeTar}
\address{
Department of Physics, University of Utah Salt Lake City, 
     UT 84112, USA
}
\author{
Olaf Kaczmarek, Frithjof Karsch and Edwin Laermann
}
\address{Fakult\"at f\"ur Physik, Universit\"at Bielefeld, 
     D-33615 Bielefeld, Germany}
\date{\today}
\maketitle
\begin{abstract}
The separation of a heavy quark and antiquark pair leads to the
formation of a tube of flux, or string, which should break in the presence
of light quark-antiquark pairs. This expected zero temperature
phenomenon has proven elusive in simulations of lattice QCD.
We present simulation results that show that the string does
break in the confining phase at nonzero temperature. 
\end{abstract}
%
%

In the absence of light quarks the heavy quark-antiquark
potential is known quite accurately from numerical simulations
of lattice quantum chromodynamics \cite{quenchedpot}. 
At large separation $R$,
the potential rises linearly, as expected in a confining
theory. The potential in the presence of light quarks
is less well determined because of the substantially
higher computational expense. Still,
all the existing lattice data at zero temperature 
\cite{fullpot1,fullpot2} agree in that they do not
show any indication of string breaking which would be
signalled by a tendency 
of the potential to level off at
large distances. The distances covered so far extend 
up to $R \lsim 2$ fm while 
it has been proposed that the dissociation threshold
would 
be reached at separations somewhere 
between 1.5 and 1.8 fm \cite{fullpot2,diss}.

In this communication we present simulation results which show
string breaking. These results have been obtained at
nonzero temperature in the confining phase of QCD.
Our work confirms trends found 
from simulations at nonzero temperature
on smaller lattices \cite{Wien}.
We have simulated QCD with two light flavours 
of staggered dynamical 
quarks on lattices of size $16^3 \times 4$
(new work) and $12^3 \times 6$
(configurations from Ref.~\cite{eos6})
at fixed values for the quark mass of
$m_q/T = 0.15$ and $0.075$ respectively. The couplings were
chosen 
to cover temperatures $T$ below the critical temperature
$T_c$ in the range of approximately
$0.7 T_c < T < T_c$. The (temperature-dependent) 
heavy quark potential $V(R,T)$ was extracted
from Polyakov loop correlations
\begin{equation}
\langle L({\vec 0}) L^\dagger ({\vec R}) \rangle = c \, 
\exp\{ - V(|{\vec R}|,T)/T\}
\end{equation}
where 
\begin{equation}
L({\vec x})= \frac{1}{3} \rm{tr} \prod_{\tau = 0}^{N_\tau - 1}
U_0({\vec x},\tau)
\end{equation}
denotes the Polyakov loop at spatial coordinates ${\vec x}$.
In the limit $R \rightarrow \infty$ the correlation function
should approach the cluster value
$|\langle L(0)\rangle |^2$ which vanishes if the potential
is rising at large distances (confinement) 
and which acquires a small
nonzero value if the string breaks.

In Figures \ref{fig:nt4} and \ref{fig:nt6} we present our
data for the potential, at the values of $\beta$ analyzed,
in lattice units. The critical couplings $\beta_c$ have been
determined as $5.306$ for $N_\tau = 4$ and $5.415$ 
for $N_\tau = 6$ respectively. The Polyakov loop correlations
have been computed not only for on-axis separations but
also for a couple of off-axis distance vectors ${\vec R}$.
Rotational invariance is reasonably well recovered if one
uses the lattice Coulomb behaviour to determine the quark-antiquark
separation, $|{\vec R}| = 1/G_{\rm lat}({\vec R})$.
The data in Figures \ref{fig:nt4} and \ref{fig:nt6} quite
clearly show a flattening of the potential at lattice
distances of about 3 to 4 lattice spacings, depending
on $\beta$.
Moreover, the height of the potential at these distances is
in nice agreement already with the infinite distance, cluster value,
shown as the right-most data point in each of the plots.

In order to obtain a rough estimate of the corresponding
temperatures in units of the critical temperature we applied
the following procedure: at the given $\beta$
and $m_q a$ values an 
interpolation formula \cite{MILCinter}
was utilized to estimate the vector meson mass $m_V a$ in lattice
units as well as the ratio of pseudoscalar to vector meson mass,
$m_{PS}/m_V$. By means of a phenomenological formula which
interpolates between the (experimentally measured) 
$\rho$ and $K^*$ mass as function
of the ratio $m_{PS}/m_V$, a 
value for $m_V$ in physical units at the simulation quark mass
can be obtained.
This number is then used to estimate the value for the lattice
spacing. This is certainly a rather crude 
procedure, yet, the resulting values for $T/T_c = a_c/a$ 
show quite stable behaviour under variations of the procedure.
The resulting temperature ratios
are summarized in table \ref{tab:temp}.

Finally, in order to facilitate a comparison 
of the $N_\tau = 4$ and $6$ results with each other and
with quenched data, the absolute scale was determined from a 
conventional Wilson loop measurement
of the string tension at zero temperature at the critical
$\beta_c$ values. The Wilson loops did not show
string breaking at the separations which could be explored.
The results for the critical temperature in units of the
string tension are obtained as
$T_c/\sqrt{\sigma} = 0.436(8)$ 
for $N_\tau = 4$ and 
$T_c/\sqrt{\sigma} = 0.462(9)$ 
for $N_\tau = 6$ \cite{MILCsigma}. 
There is a substantial (25\%) difference between
scales set by $m_V$ and $\sqrt{\sigma}$, which suggests a
magnitude for systematic errors in the scale estimate.

In Figure \ref{fig:norm} we show the potential in the
presence of dynamical quarks in physical units. The data
has been normalized to the cluster value i.e. the self
energies have been taken out. The potential is flat within
the error bars at distances larger than about 1 fm. It also
seems that the turn-over point is slightly $T$ dependent,
becoming smaller with increasing temperature. It is beyond
the scope of the quality of the data at this stage, however,
to quantify this statement.

Assuming that the Wilson loop string tension is not
affected by the absence of dynamical fermions one can
then immediately compare quenched and full QCD potentials
in physical units at the same temperature, 
as is shown in Fig.~\ref{fig:phys}. The quenched data
has been taken from \cite{Olaf} and was obtained in the same
way, i.e. computed from Polyakov loop correlations. 
Each data set has slightly been shifted up or down to 
give rough agreement at intermediate distances 
around 0.3 fm. 
Figure~\ref{fig:phys} contains, for further
comparison, the dashed line denoting
$-\pi/(12 R) + (420 {\rm MeV})^2 R$ 
which gives a good description of
the zero temperature quenched potential.
Note that the nonzero-temperature
quenched potential is rising with distance $R$ but the slope 
decreases with temperature, i.e. the (quenched)
string tension is temperature 
dependent and becomes smaller closer to the critical $T_c$. 
Again, the comparison with quenched potentials at the same
temperature demonstrates quite nicely that
the potential in the presence of dynamical quarks becomes 
flat within the error bars at distances of about 1 fm.
From Figure~\ref{fig:phys} we conclude that the observed
string breaking, albeit at nonzero temperature, is an effect
caused by the presence of dynamical fermions. 

We have seen that string breaking is relatively easy to observe in the
Polyakov loop correlation, while it is difficult to detect through the
conventional Wilson loop observable.  Why is this so?  The Wilson loop
observable creates a static quark-antiquark pair together with a flux
tube joining them.  In the presence of such a static pair at large
$R$, we expect the correct ground state of the Hamiltonian to consist
of two isolated heavy-light mesons, however.  Such a state with an
extra light dynamical quark pair has poor overlap with the flux-tube
state, so it is presumably revealed only after evolution to a very
large $T$.  An improved Wilson-loop-style determination of the heavy
quark potential in full QCD would employ a variational superposition
of the flux-tube and two-heavy-meson states\cite{drummond,Knechtli}. The
Polyakov loop approach, on the other hand, although limited in
practical application to temperatures close to or above $T_c$, 
builds in no
prejudices about the structure of the static-pair ground state wave
function.  Screening from light quarks in the thermal ensemble occurs
readily.


This work was supported by the TMR network ERBFMRX-CT-970122 and 
the NATO-CRG 940451. 
C.D. gratefully acknowledges support from the US National Science
Foundation and the Zentrum f\"ur Interdisziplin\"are Forschung,
Universit\"at Bielefeld, where this work was initiated.
We thank the MILC collaboration for use of previously unpublished
data.  MILC collaboration calculations were carried out on the
following: the Intel Paragon at Indiana University, the IBM SP2 at the
Cornell Theory Center, the IBM SP2 at the University of Utah, and the
workstation cluster at SCRI, Florida State University.



\begin{table}
\caption{Estimates of the temperature at the various $\beta$
         and $N_\tau$ values.}
\label{tab:temp}
\begin{tabular}{llcccc}
$N_\tau = 4$ & $\beta$ & 5.10 & 5.20 & 5.25 & 5.28 \\
\cline{2-6}
             & $T/T_c$ & 0.67 & 0.79 & 0.87 & 0.94 \\
\hline
$N_\tau = 6$ & $\beta$ & 5.37 & 5.38 & 5.39 & 5.41 \\
\cline{2-6}
             & $T/T_c$ & 0.85 & 0.88 & 0.91 & 0.98 \\
\end{tabular}
\end{table}

\figure{
 \epsfig{bbllx=100,bblly=330,bburx=530,bbury=740,clip=,
         file=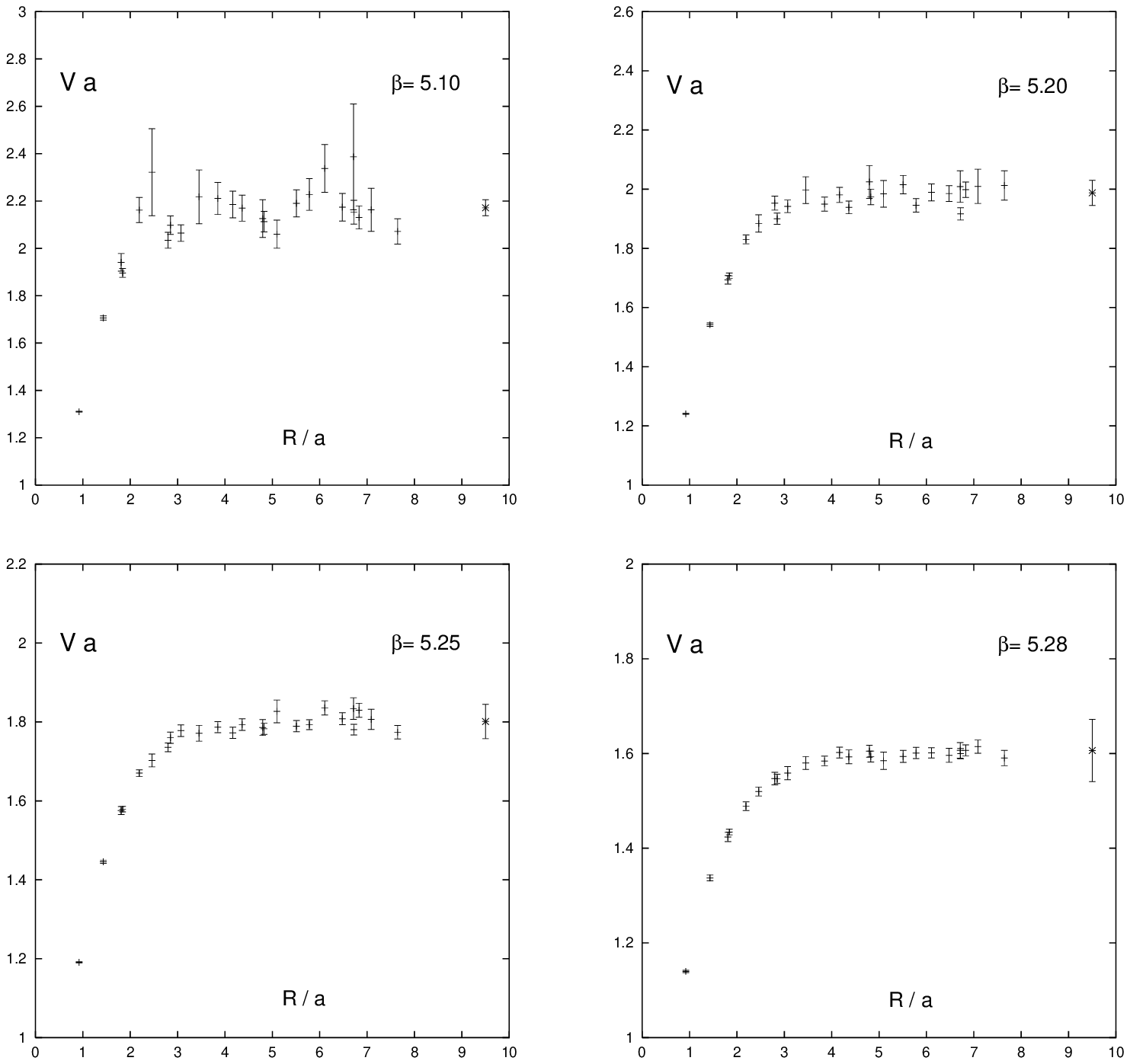,width=154mm}
\caption{The potentials in lattice units at the $\beta$ values
         analyzed for $N_\tau=4$. The right-most data points 
         plotted at $R/a = 9.5$ and denoted by stars 
         are the infinite distance,
         cluster values $-T {\rm ln} |\langle L \rangle |^2$.
         }
\label{fig:nt4}
}
\figure{
 \epsfig{bbllx=110,bblly=280,bburx=555,bbury=710,
         file=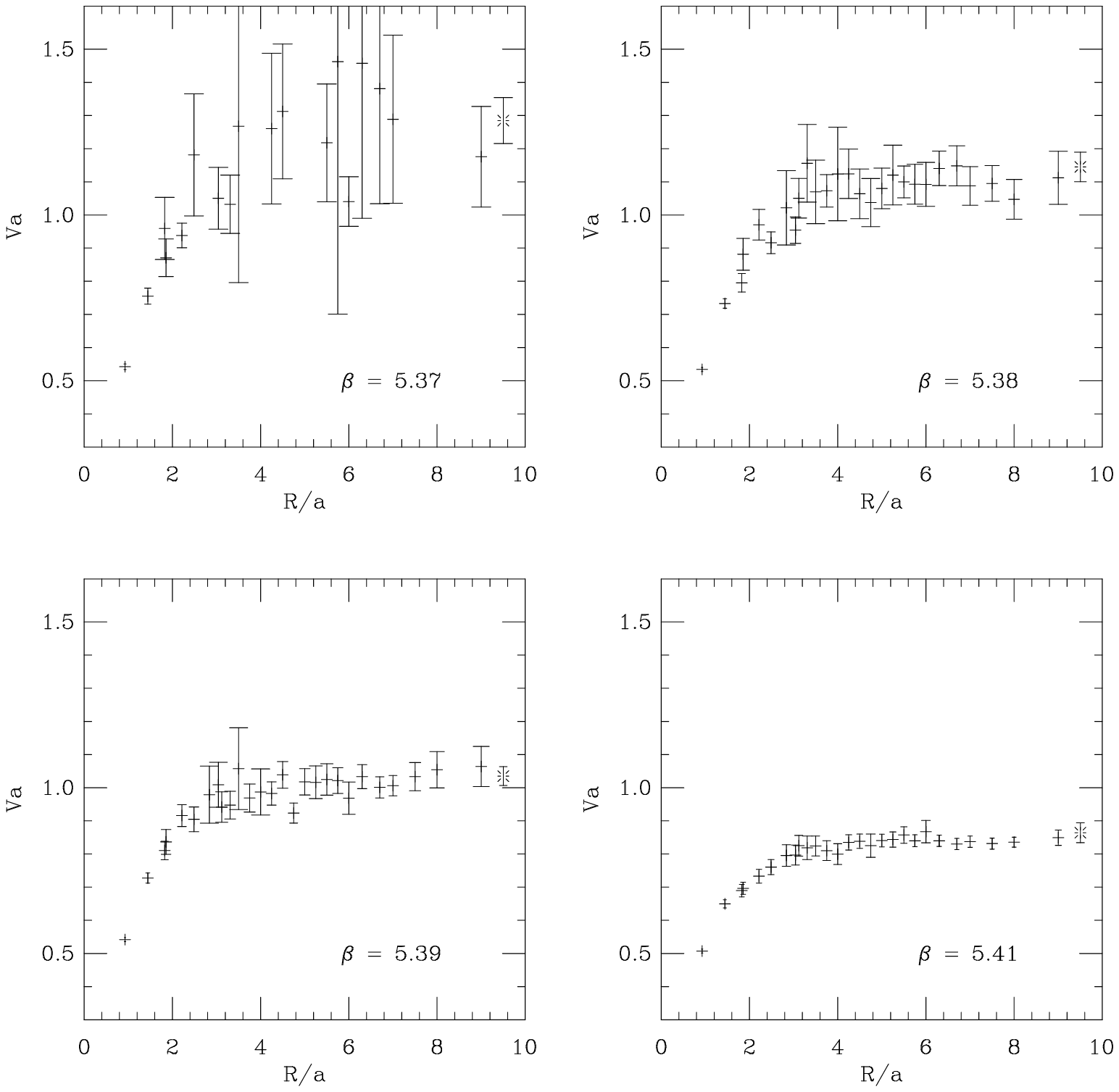,width=154mm}
\caption{The potentials in lattice units at the $\beta$ values
         analyzed for $N_\tau=6$. The right-most data points 
         plotted at $R/a = 9.5$ and denoted by stars 
         are the infinite distance,
         cluster values $-T {\rm ln} |\langle L \rangle |^2$.
         }
\label{fig:nt6}
}
\figure{
 \epsfig{file=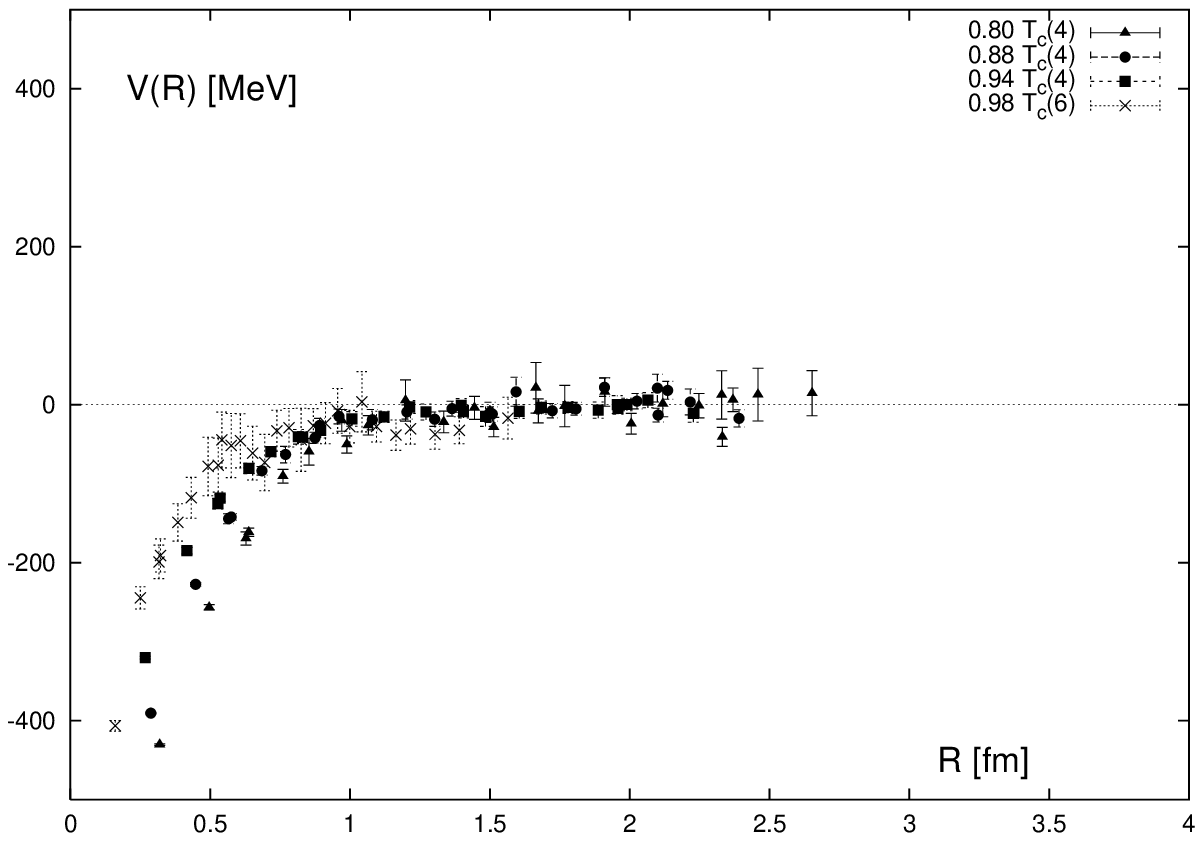,width=154mm}
\caption{The potential in physical units at various temperatures.
         The results are from lattices with $N_\tau = 4$ and $6$, 
         indicated by the number in brackets. The data has been
         normalized to the cluster value.
}
\label{fig:norm}
}
\figure{
 \epsfig{file=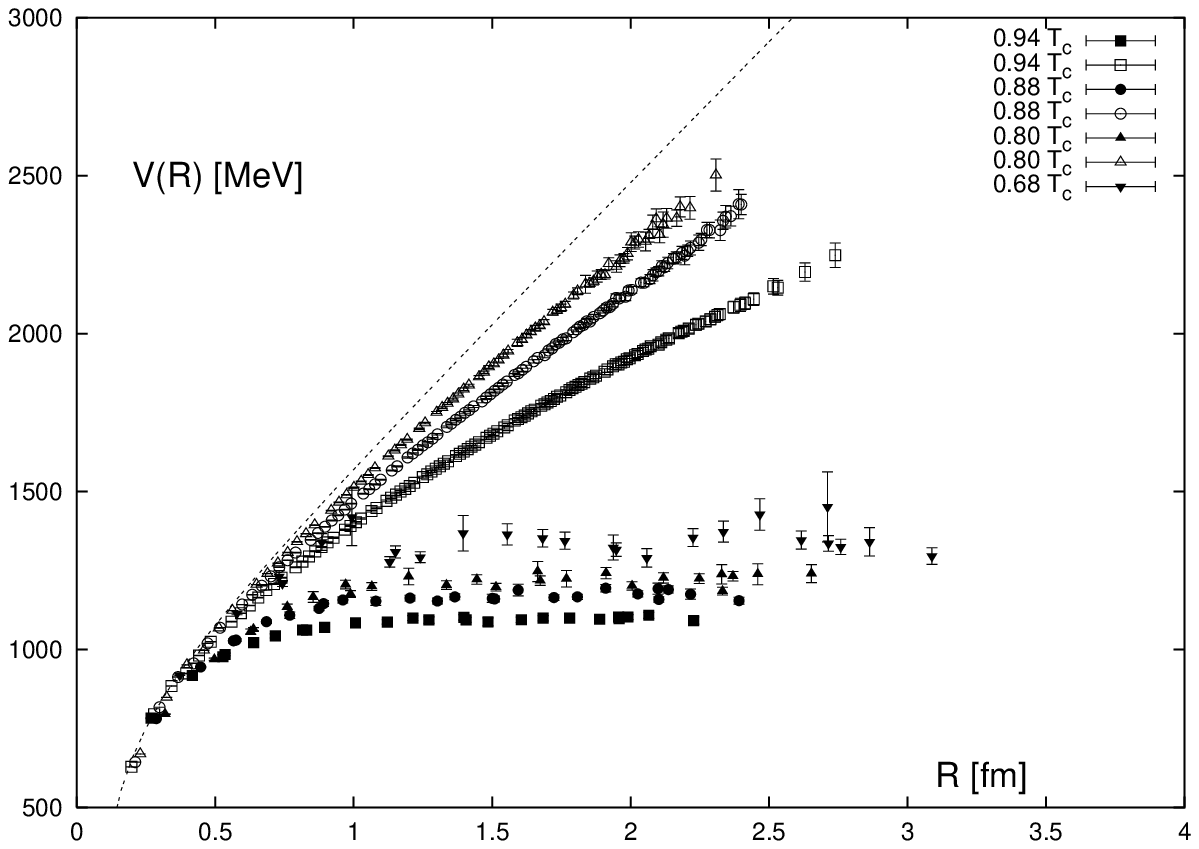,width=154mm}
\caption{The potential in physical units at various temperatures.
         Compared are quenched (open symbols) and 
         full (filled symbols) QCD potentials at the same 
         temperature. The dashed line is the zero temperature
         quenched potential. 
         The data has been normalized to
         agree at distances around 0.3 fm.
}
\label{fig:phys}
}

\end{document}